\documentclass[aps,prl,twocolumn,groupedaddress]{revtex4}

\usepackage{graphicx}

\begin{document}


  \title{Detection and Imaging of He$_2$ Molecules in Superfluid Helium}


\author{W. G. Rellergert, S. B. Cahn, A. Garvan, J. C. Hanson,
  W. H. Lippincott, J. A. Nikkel \& D. N. McKinsey}
\email[Electronic address:  ]{daniel.mckinsey@yale.edu}
\affiliation{Department of Physics, Yale University, New Haven, Connecticut 06520, USA}


\date{\today}

\begin{abstract}

We present data that show a cycling
transition can be used to detect and image metastable He$_2$ triplet molecules in
superfluid helium.
We demonstrate that limitations on the cycling efficiency due to the vibrational
structure of the molecule can be mitigated by the use of repumping lasers.
Images of the molecules obtained using the method are also shown.  
This technique gives rise to a new kind of ionizing radiation detector.  The use of He$_2$ triplet molecules as
tracer particles in the
superfluid promises to be a powerful tool for visualization of both
quantum and classical turbulence in liquid
helium.
\end{abstract}


\maketitle


Ionizing radiation events in liquid helium produce unstable He$_2$ molecules
in both singlet and triplet states~\cite{Fleishman1959,Surko1968,Dennis1969,Hill1971}.
The singlet state molecules radiatively decay in
a few nanoseconds~\cite{Hill1989}, but the triplet state molecules are metastable because a radiative transition to the
ground state of two free atoms requires a strongly forbidden spin flip.
The radiative lifetime of the triplet molecules has been calculated to be 18 s in vacuum~\cite{Chabalowski1989} and
measured to be 13 s in liquid helium~\cite{McKinsey1999}.  Here we present
data supporting our previous proposal~\cite{McKinsey2005} to detect and image
the triplet molecules by driving them
through multiple fluorescence-emitting transitions during their lifetime.

The lowest-lying electronic states and two relevant vibrational
levels of the triplet molecules are shown in Fig. 1, as well as one cycling
transition used to
detect them.
Two infrared photons can excite a triplet molecule from the ground
$\mathit{a}^{3}\Sigma^{+}_{\mathit{u}}$ state to the $\mathit{d}^{3}\Sigma^{+}_{\mathit{u}}$ state.
Calculations of the branching
ratios indicate that about 10\% of the excited molecules will decay to the
$\rm \mathit{c}^{3}\Sigma^{+}_{\mathit{g}}$ state, while the
remaining 90\% will decay to the $\rm \mathit{b}^{3}\Pi_{\mathit{g}}$ state,
emitting detectable red photons at 640 nm.
Molecules in both the $\rm \mathit{c}^{3}\Sigma^{+}_{\mathit{g}}$ and $\rm
\mathit{b}^{3}\Pi_{\mathit{g}}$ states then decay back to the
$\mathit{a}^{3}\Sigma^{+}_{\mathit{u}}$ state, and the
process can be repeated.  Since the $\rm \mathit{d}^{3}\Sigma^{+}_{\mathit{u}}
\rightarrow \rm \mathit{b}^{3}\Pi_{\mathit{g}}$ transition emits a photon that
is well separated in wavelength from the excitation
photons, scattered laser
light can be blocked by appropriate filters.
As the entire cycle occurs in roughly 50 ns, it could in principle be repeated
enough times to allow for single molecule detection.

\begin{figure}
\includegraphics[width=2.5 in]{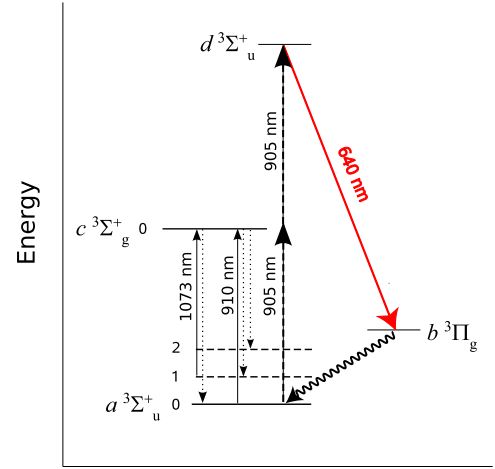}
\centering
\caption{Shown schematically are the lowest energy levels of the He$_2$ triplet molecule in
  liquid  helium.  A cycling transition used to detect and image the
  molecules is indicated, as well as transitions that control the populations of the
  $\mathit{a}^{3}\Sigma^{+}_{\mathit{u}}$ vibrational levels.}
\label{fig:NatureFIG1pics} 
\end{figure}

The molecular structure complicates matters because molecules in an excited electronic
 state may decay to excited rotational or vibrational levels of the
electronic ground state.  If those levels have long relaxation times, and are out of resonance with the
excitation lasers, the rate at which a molecule can be
cycled is greatly reduced.  For He$_2$ molecules in the
liquid, the absorption spectral lines are ~120 cm$^{-1}$ wide~\cite{Hill1971,Eltsov1995}
which is considerably larger than the spacing of the rotational levels (7 cm$^{-1}$)~\cite{Herzberg1983}. 
The vibrational levels, on the other hand, are
separated by about 1500 cm$^{-1}$~\cite{Herzberg1983}, and the vibrational relaxation time is over
100 ms~\cite{Eltsov1995}.
Therefore, molecules falling to excited vibrational levels of the $\mathit{a}^{3}\Sigma^{+}_{\mathit{u}}$
state are trapped in off-resonant levels and are lost for subsequent cycles.
They can be
recovered, however, with the use of repumping lasers. 
As an example, a molecule that decays to the first vibrational level of
the $\mathit{a}^{3}\Sigma^{+}_{\mathit{u}}$ state, $\it{a}$(1), can be driven into
the zeroth vibrational level of the $\rm
\mathit{c}^{3}\Sigma^{+}_{\mathit{g}}$ state, $\it{c}$(0), with light at 1073 nm
(Fig. 1).  
Our calculations of the Franck-Condon factors imply that a molecule in $\it{c}$(0)
will decay back to $\it{a}$(0) around 95\% of the time. 
As there are only two
excited vibrational levels  below $\it{b}$(0), two repumping lasers are
likely all that are needed to ensure that molecules are recovered for high
cycling rates in this scheme.

A typical 1 MeV electronic recoil event in liquid helium creates about 32,000 He$_2$ molecules.
About 60\% of those molecules are
created in the 
singlet state and 40\% are created in the triplet state~\cite{Adams2001}.
For our laser-induced fluorescence studies, we used a 1 $\mu$Ci $^{113}$Sn beta
source immersed in the superfluid (1.9 $\pm$ 0.1 K) to create the molecules.  The 364 keV betas emitted by the
$^{113}$Sn nuclei deposit most of their energy within 1 cm~\cite{Adams2001} of the source and
create about 2.5$\times$10$^{8}$ triplet molecules in steady state.
Laser beams illuminate this region, and the resulting  fluorescence is
recorded with either a
photomultiplier tube or a camera as shown in Fig. 2. 

\begin{figure}
\includegraphics[width=2.5 in]{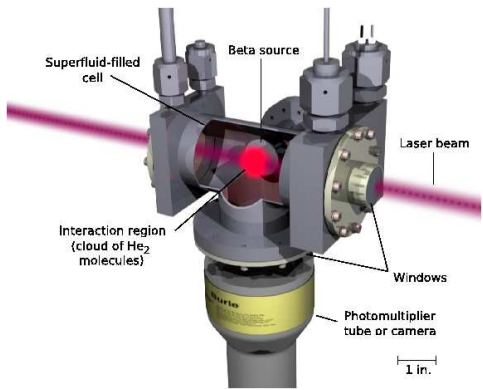}  
\centering
\caption{Experimental setup.  Lasers enter and exit the superfluid-filled cell
through 25 mm diameter windows and a light detector views the interaction
region through a 25 mm diameter window on the bottom of the cell.  The cut-away section
in the middle shows the radioactive beta
source and the interaction region.  The source is deposited on a 25 mm diameter metal
disk and is mounted on a flange on the side of the cell.  It
can also be mounted from the top of the cell.}
\end{figure}

Figure 3a shows excitation spectra we obtain by scanning a tunable Q-switched pulsed laser \cite{laser} in combination with
a continuous-wave diode laser.  The wavelength of the pulsed laser is
scanned from 760 nm to 1200 nm in steps of 1 nm.  The pulse energy is 5 mJ and the repetition rate is 2 Hz.  
The pulsed laser beam is overlapped with the diode laser beam and both are
expanded to a spot size of 1 cm$^2$.  The solid red curve
shows the fluorescence signal as the pulsed laser is scanned with a 60 mW/cm$^2$ diode
laser at 1073 nm.  The filters used to block the
scattered excitation light allow for the detection of fluorescence in the wavelength
range from 515 nm to 750 nm.  Peaks in the fluorescence signal are observed
when the pulsed laser wavelength is at either 800 nm or 905 nm.  The wavelength
dependence of the fluorescence is measured with a
monochromator equipped with short-pass filters that transmit wavelengths less than 875 nm.
For both 800 nm and 905 nm excitation, all observed fluorescence is contained in one peak centered at 640 nm which
is consistent with $\rm \mathit{d}^{3}\Sigma^{+}_{\mathit{u}} \rightarrow \rm
\mathit{b}^{3}\Pi_{\mathit{g}}$ emission (Fig. 3b).  It should be noted that
the wavelength resolution is limited by the slit size of the
monochromator.  A fit of the average
fluorescence signal detected by the photomultiplier tube yields 48 $\pm$ 2 ns for the
$\rm \mathit{d}^{3}\Sigma^{+}_{\mathit{u}}$ state lifetime (Fig. 3c).

\begin{figure}
\includegraphics[width=3.375in]{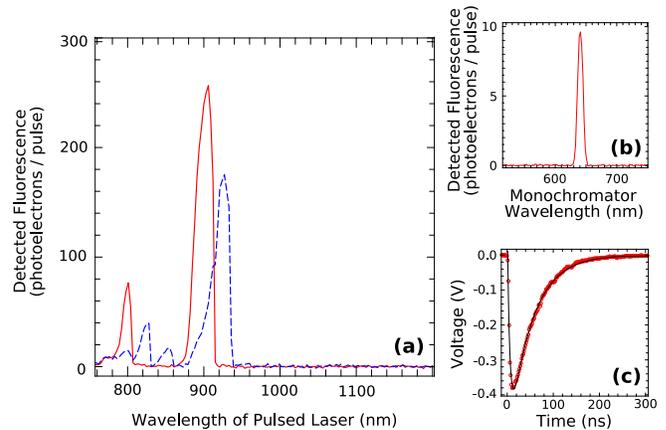} 
\centering
\caption{Laser-induced fluorescence data. (a) Two
  excitation spectra
obtained by scanning a pulsed laser from 760 nm to 1200 nm in combination
with a diode laser are shown on the
left.  The solid red line shows the fluorescence signal intensity when a diode laser
at 1073 nm is used.  The dashed blue line shows the signal when
a diode laser at 910 nm is substituted for the 1073 nm diode laser.  (b)  The wavelength dependence
of the fluorescence resulting from laser pulses at 905 nm is measured using a monochromator.  
All observed fluorescence is centered at 640 nm.  (c) The
red circles show the average trace from a photomultiplier tube
when the pulsed laser is fixed at 905 nm.  The black line shows a fit
which gives a 48 $\pm$ 2 ns lifetime for the $\rm
\mathit{d}^{3}\Sigma^{+}_{\mathit{u}}$ fluorescence.}
\end{figure} 

The
fluorescence resulting from
laser pulses near 800 nm was previously observed by Benderskii \textit{et
al}~\cite{Benderskii1999,Benderskii2002}.  By calculating potential energy curves for molecules
in liquid helium, they determined that the $\rm \mathit{d}^{3}\Sigma^{+}_{\mathit{u}}$
state is shifted up in energy when the molecule is in the liquid.  The shift
is attributed to the change in size of the bubble formed by the molecule.  
The two-photon absorption line at 800 nm is therefore thought to drive $\it{a}$(0) $\rightarrow$ $\it{c}$(1) $\rightarrow$ $\it{d}$(2).
As the difference in energy between two 905 nm photons and two 800 nm photons is
3000 cm$^{-1}$, which is roughly twice the expected spacing for the
vibrational levels of the $\rm \mathit{d}^{3}\Sigma^{+}_{\mathit{u}}$ state~\cite{Herzberg1983},
it is likely that the peak at 905 nm is driving $\it{a}$(0) $\rightarrow$
$\it{c}$(0) $\rightarrow$ $\it{d}$(0).  As
our attempts to produce more fluorescence by using two pulsed lasers at different
wavelengths were unsuccessful, a single pulse at 905 nm appears
to be the most efficient way to excite the molecules to $\it{d}$(0).  Being able
to excite the molecules with only one pulsed laser, instead of two as originally proposed~\cite{McKinsey2005},
is a major simplification.  With single wavelength
excitation, 2D position information can be obtained by either rastering the
laser beam, or taking an image with a camera.  A second scan (or image) in an
orthogonal direction can then yield 3D information.  

The dashed blue curve in Fig. 3a shows the resulting fluorescence when a 90 mW/cm$^2$ diode
laser at 910 nm is substituted for the diode laser at 1073 nm.
Shown schematically in Fig. 1, this laser depletes $\it{a}$(0) by driving
molecules into $\it{a}$(1)
and $\it{a}$(2), which are long lived.
The peaks at 800 nm and 905 nm are significantly reduced
showing that the population  of $\it{a}$(0) is indeed lowered, and new peaks in
fluorescence are observed at 825 nm, 850 nm, and 925 nm.  The suspected
excitation scheme for these peaks is $\it{a}$(1) $\rightarrow$ $\it{c}$(2) $\rightarrow$ $\it{d}$(3) for
825  nm, $\it{a}$(2) $\rightarrow$ $\it{c}$(3) $\rightarrow$ $\it{d}$(4) for 850 nm, and $\it{a}$(1) $\rightarrow$ $\it{c}$(1) $\rightarrow$ $\it{d}$(1) for 925 nm.    

The ability to repump molecules that fall to $\it{a}$(1) is demonstrated in
Fig. 4, which
shows the dependence of the fluorescence signal on the pulse number in a
sequence of 64 consecutive  pulses of 905 nm laser light (2.5 mJ/cm$^2$).  Before
the pulse sequence, the laser pulses are blocked with a shutter for 6.4 s to allow the
molecules to come to equilibrium.  The shutter then opens, and the molecules
are excited with a laser pulse every 100 ms for 6.4 s.  The shutter then
closes, and the process is repeated.  Figure 4 shows the average of ten such
consecutive pulse sequences both with and without the use of a repumping diode laser.
Without a repumping laser,
the signal drops to 25\% of the initial value by the 
tenth pulse indicating that molecules are being
lost by some mechanism when cycled.  With the addition of a
diode laser at 1073 nm, however, the signal remains above 85\% of the
initial value.  These data demonstrate that the molecules are indeed
being cycled multiple times, that the primary loss mechanism is
de-excitations to $\it{a}$(1), and that the molecules in $\it{a}$(1) can be
driven back to $\it{a}$(0) by a repumping laser.  

\begin{figure}
\includegraphics[width=2.5 in]{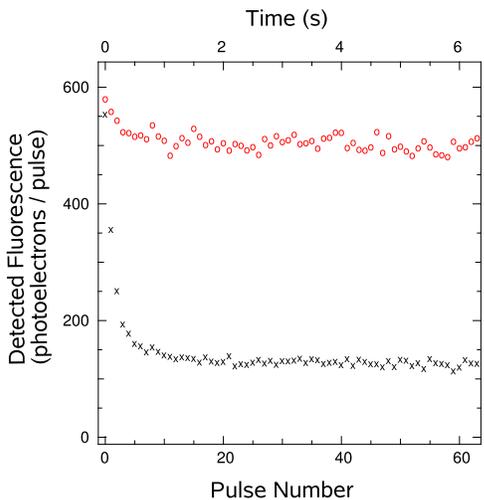}
\centering
\caption{These data
show the dependence of the fluorescence signal on the pulse number in a
sequence of 64 consecutive pulses of 905 nm laser light.  
Without a repumping laser (black x's),
the signal drops to 25\% of the initial value by the tenth pulse.  With the addition of a
diode laser at 1073 nm (red circles), however, the signal remains above 85\% of the
initial value.
} 
\end{figure} 

The size of the signal in Fig. 4 can be used to obtain a lower bound on the percentage of molecules
that are emitting a 640 nm photon.  Taking into account the laser beam size
and its position, we estimate that it overlaps at most 30\% of the molecules
produced by the beta source.
The solid angle subtended by the photomultiplier tube (0.75\%), its quantum efficiency (5\%), and the transmission
of the various windows and optical filters (30\%) indicate that about 1 out of every 10$^4$ emitted
photons are detected.  Therefore, detection of more than 500 photoelectrons suggests
that for each pulse at least 5\% of the molecules in the laser beam are emitting a 640 nm
photon.

Images of the molecules are obtained with amplified
CCD cameras.  A false-color image of the cloud of molecules created
by the beta source is shown in Fig. 5a.   The fluorescence is produced by pulsing one laser at 905 nm and
a second laser at 925 nm, exciting molecules in both $\it{a}$(0) and
$\it{a}$(1).  The repetition rate of each laser is 10 Hz, and the pulse energies are
5 mJ. The total exposure time is 5 minutes.
The spatial extent of the cloud agrees with the expected beta particle track
length of 1 cm taking into account the 5 mm distance from the laser beam to the source.
An image taken with a collimator on the beta
source is shown in Fig. 5b.  A single pulsed laser at
905 nm and a  60 mW/cm$^2$ continuous-wave repumping laser at 1073 nm
are used to produce the fluorescence.  The repetition rate and pulse energy
are again 10 Hz and 5 mJ, respectively, but the
exposure time is 2.5 minutes. 
As expected, the shape of the cloud is
determined by the limited angular spread of the beta particle trajectories.

\begin{figure}
\includegraphics[width=2.5 in]{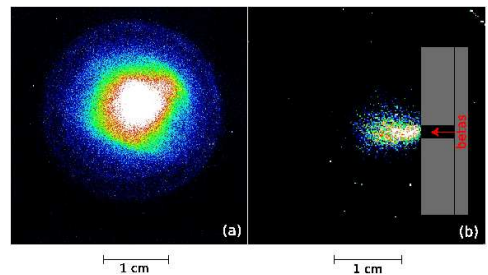} 
\centering
\caption{Images of the molecules
  obtained using amplified CCD cameras. (a) In the image on the
left, the beta source is facing the camera.     
(b)  The image on the right is formed by collimating the beta source
and imaging it from the side.  A sketch of the source and collimator has been
added to show its location.  
}
\end{figure}

The technique outlined above gives rise to a new detector technology using
liquid helium.  In addition to the qualities and applications of such a
detector that have already been
proposed~\cite{McKinsey2005}, the method may be useful for a Compton gamma
ray imager.  If a gamma ray scatters in liquid helium and is subsequently absorbed in a second detector, the locations
of those events and total energy deposited can be used to kinematically determine the energy and
trajectory of the incoming gamma ray.  Helium has three characteristics that
are desirable for such a detector.  First, the low atomic number of helium
makes it more likely that a gamma ray will scatter in the liquid rather than be
absorbed.  Second, the uncertainty in the initial momentum of the
electron from which a gamma ray scatters is smaller 
than it would be for materials with higher atomic numbers, leading to a
lower uncertainty in the determination of the initial energy and direction of
a scattered gamma ray.  Finally, the track of
molecules produced by the recoiling Compton electron may be imaged to further reduce the uncertainty
in the initial gamma ray direction.

Laser-induced fluorescence of the molecules also provides a new tool for
visualization of quantum and classical turbulence in liquid helium~\cite{Niemela2006,Charalambous2006,Vinen2006}.
Particle image velocimetry has been used to visualize liquid helium flows~\cite{Zhang2005,White2002}, but it appears
that the required micron-sized particles are too large to act as passive
tracers.  A possible reason is 
their interaction with vortices in the liquid.  The He$_2$ triplet molecules are much smaller (7
\AA \ radius~\cite{Benderskii2002}), and
should be unaffected by vortices at temperatures above 1 K~\cite{Vinen2006a}.
Also, the
small size of the molecules might allow for resolution of the Kolmogorov length
scale.  Additionally, at sufficiently low temperatures the molecules will become trapped on
vortices~\cite{Vinen2006a} as has been observed for
ions~\cite{Yarmchuk1979,Guo2007} and micron-sized solid hydrogen
particles~\cite{Bewley2006}.  Stereoscopic imaging of the
laser-induced fluorescence of the molecules could then be used to obtain
3D images of vortex lines and their dynamics in the superfluid.    

Seeding the flow with the triplet molecules can be done
in a relatively easy and unintrusive manner with a radioactive source or an
intense, focused laser pulse.  By adding a heater in the liquid helium, and
monitoring the distribution of the triplet
molecules as a function of heater power, one can obtain their diffusion
coefficient in the same way that neutron absorption tomography allowed for the
determination of the diffusion coefficient of $^3$He~\cite{Lamoreaux2002}.  That information also
allows one to obtain velocity fields in liquid
helium~\cite{Hayden2004}.  

In addition, the vibrational levels of the triplet molecule can be used to image and tag a
group of molecules
in a specific region of the liquid and image their location some time later.
For example, after pumping all molecules into $\it{a}$(0), a laser pulse
at 800 nm or 905 nm will result in $\rm
\mathit{d}^{3}\Sigma^{+}_{\mathit{u}} \rightarrow \rm
\mathit{b}^{3}\Pi_{\mathit{g}}$ fluorescence and also create a
population in $\it{a}$(1) which would otherwise not be
present.  A well collimated laser pulse at one of those wavelengths can therefore be used to image a line of
molecules, and, a short while
later, a second, expanded laser pulse at 925 nm would show how that line has deformed by
imaging those molecules that fell to $\it{a}$(1) after the first
excitation.  This method is similar to that of a study that
used oxygen molecules to measure turbulence in air, and allows one to measure
transverse velocity increments which can be used to determine the single-point probability
density of velocity fluctuations~\cite{Noullez1997}.  



In conclusion, we have detected and imaged He$_2$
triplet molecules in liquid helium using laser-induced fluorescence.  We have demonstrated good
control over the vibrational structure of the molecules with the use of
continuous-wave diode lasers which allow the molecules to be cycled
multiple times over the course of their 13 s lifetime.  The cycling rate has
so far been limited only by the 10 Hz repetition
rate of our lasers.  

\begin{acknowledgments}
This work was supported by the Defense Threat Reduction Agency under grant DTRA01-03-D-0009-0011.
\end{acknowledgments}




\end{document}